\documentclass[traditabstract]{aa}
\usepackage{longtable}
\usepackage{natbib,afterpage}
\usepackage{graphicx,color}
\usepackage{amssymb,times,graphics, subfigure}
\usepackage[english]{babel}
\usepackage{txfonts}
\usepackage{subfigure}

\def\Msun{\hbox{M$_{\odot}$}}               
\def\arcsec{\hbox{$^{\prime\prime}$}}
\def\arcmin{\hbox{$^{\prime}$}}

\def\deg{\hbox{$^\circ$}}       

\bibpunct{(}{)}{;}{a}{}{,}

\begin{document}
\selectlanguage{english}
\newcommand{\red}{\textcolor[rgb]{1,0,0}}
  \title{Discovery of multiple dust shells beyond 1\,arcmin in the circumstellar envelope of IRC\,+10216  using Herschel/PACS
\thanks{Herschel is an ESA space observatory with science instruments provided by European-led Principal Investigator consortia and with important participation from NASA. 
}}

 \author{L. Decin\inst{1,2}
  \and
  P. Royer\inst{1}
  \and
  N.L.J. Cox\inst{1}
  \and
  B. Vandenbussche\inst{1}
  \and
  R. Ottensamer\inst{3}
  \and	
  J.A.D.L. Blommaert\inst{1}
  \and
  M.A.T. Groenewegen\inst{4}
  \and	
  M.J. Barlow\inst{5}
  \and
  T. Lim\inst{6}
  \and
  F. Kerschbaum\inst{3}
  \and	
  T. Posch\inst{3}
  \and
  C. Waelkens\inst{1}
}

  \offprints{Leen.Decin@ster.kuleuven.be}

  \institute{
Instituut voor Sterrenkunde,
             Katholieke Universiteit Leuven, Celestijnenlaan 200D, 3001 Leuven, Belgium
 		\and
	Sterrenkundig Instituut Anton Pannekoek, University of Amsterdam, Science Park 904, NL-1098 Amsterdam, The Netherlands
    \and
 University of Vienna, Department of Astronomy, T{\"u}rkenschanzstra\ss{}e 17, A-1180 Vienna, Austria
  \and
Royal Observatory of Belgium, Ringlaan 3, B-1180 Brussels, Belgium 
    \and
 Dept.\ of Physics \& Astronomy, University College London, Gower St, London WC1E 6BT, UK 
  \and
  Space Science and Technology Department, Rutherford Appleton Laboratory, Oxfordshire, OX11 0QX, UK
}


  \date{Received / accepted }

 \abstract{We present new Herschel/PACS images at 70, 100, and 160\,$\mu$m of the well-known, nearby, carbon-rich asymptotic giant branch star IRC+10216 revealing multiple dust shells in its circumstellar envelope. For the first time, dust shells (or arcs) are detected until 320\arcsec.  The almost spherical shells are non-concentric and have an angular extent between $\sim$40\deg\ and $\sim$200\deg. The shells have a typical width of 5\arcsec--8\arcsec, and the shell separation varies in the range of $\sim$10\arcsec--35\arcsec, corresponding to $\sim$500\,--1\,700\,yr. Local density variations within one arc are visible. The shell/intershell density contrast is typically $\sim$4, and the arcs contain some 50\% more dust mass than the smooth envelope. The observed (nested) arcs record the mass-loss history over the past 16\,000\,yr, but Rayleigh-Taylor and Kelvin-Helmholtz instabilities in the turbulent astropause and astrosheath will erase any signature of the mass-loss history for at least the first 200\,000\,yr of mass loss. Accounting for the bowshock structure, the envelope mass around IRC+10216 contains $>$2\,\Msun\ of gas and dust mass. It is argued that the origin of the shells is related to non-isotropic mass-loss events and clumpy dust formation. }
{}
{}
{}
{}

\titlerunning{Multiple dust shells beyond 1 arcmin in the CSE of IRC+10216}
  \maketitle
%

\section{Introduction} \label{Introduction}

Stars of low and intermediate initial mass lose a significant fraction of their mass  when evolving along the asymptotic giant branch (AGB). Pulsations in combination with radiation pressure on newly formed dust grains are thought to be chiefly responsible for driving the stellar wind. During the mass-loss event, a large circumstellar envelope (CSE) is formed. The mass-loss history of the stars is imprinted, in part, into the structure of the extended CSE. In the past decade, more and more evidence has been found that the envelope of many AGBs show deviations from spherical symmetry and/or exhibit temporal density variations \citep{Neri1998A&AS..130....1N, Schoier2005A&A...436..633S, Decin2007A&A...475..233D}.

IRC\,+10216 (CW~Leo) is the nearest (carbon-rich) AGB star \citep[D$\sim$150\,pc,][]{Crosas1997ApJ...483..913C} and serves as an archetype for the study of AGB stars. It has been the topic of many studies focussing on the structure of its thick envelope, which is created by a stellar wind carrying away the material at a rate of $\sim$1$\times 10^{-5}$\,\Msun/yr \citep{Decin2010Natur.467...64D}. On small spatial scales (arcsec and below), the central regions of IRC\,+10216 show the presence of a peanut- or bipolar-like structure \citep[e.g.,][]{Mauron2000A&A...359..707M,Menut2007MNRAS.376L...6M} together with a series of clumps \citep[e.g.,][]{Weigelt2002A&A...392..131W}. Despite the presence of asymmetric structure close to the star, \citet{Mauron1999A&A...349..203M}  have detected multiple, almost concentric, shells (or arcs) out to a distance of 55\arcsec\ in radius on top of the smooth extended envelope. The observed arcs are thought to be the result of ambient Galactic light scattered by the dusty envelope, of which the shell/intershell density contrast is about of three \citep{Mauron2000A&A...359..707M}. The shells are incomplete and cover $\sim$30\deg-90\deg\ in azimuth with respect to the central star.  Deep V-band images taken by \citet{Leao2006A&A...455..187L} show that the nested shells appear to extend to $\sim$80\arcsec\ and are composed of thinner elongated shells. The shells have a typical width of $\sim$1.6\arcsec\ and are separated by $\sim$5\arcsec--20\arcsec, which corresponds to $\sim$250--1000\,yr.

In this paper, we present new infrared images (at 70, 100, and 160\,$\mu$m) of the dusty envelope around IRC+10216. For the first time, multiple dust shells are detected up to a distance of 320\arcsec\ from the central star. Although limited by the sensitivity of the data, these observations suggest that the shells are present  all the way to the bowshock at $\sim$450\arcsec\ at 100\,$\mu$m \citep[see also][]{Ladjal2010A&A...518L.141L}. In Sect.~\ref{Observations}, we present the Herschel/PACS observations. We analyse, in Sect.~\ref{Results}, the morphology of the dusty envelope around IRC+10216 and the compatibility of the PACS infrared observations with optical data. We then discuss, in Sect.~\ref{Discussion}, the density variations in the arcs, the mass-loss history and the possible origin of the multiple shells. Finally, the conclusions are presented in Sect.~\ref{Conclusions}.

\section{Observations and data reduction} \label{Observations}

\begin{figure*}[!htp]
     \begin{minipage}[t]{.062\textwidth}
    \end{minipage}
     \begin{minipage}[t]{.561\textwidth}
        \centerline{\textbf{\phantom{AAAAAAAA}NOT DECONVOLVED}}
    \end{minipage}
    \hfill
    \begin{minipage}[t]{.379\textwidth}
        \centering{\textbf{DECONVOLVED}}
    \end{minipage}
    
    \begin{minipage}[t]{.062\textwidth}
        \vspace*{-4cm}
        \centering{\textbf{\large{70\,$\mu$m}}}
    \end{minipage}
   \begin{minipage}[t]{.561\textwidth}
        \centerline{\resizebox{\textwidth}{!}{\includegraphics{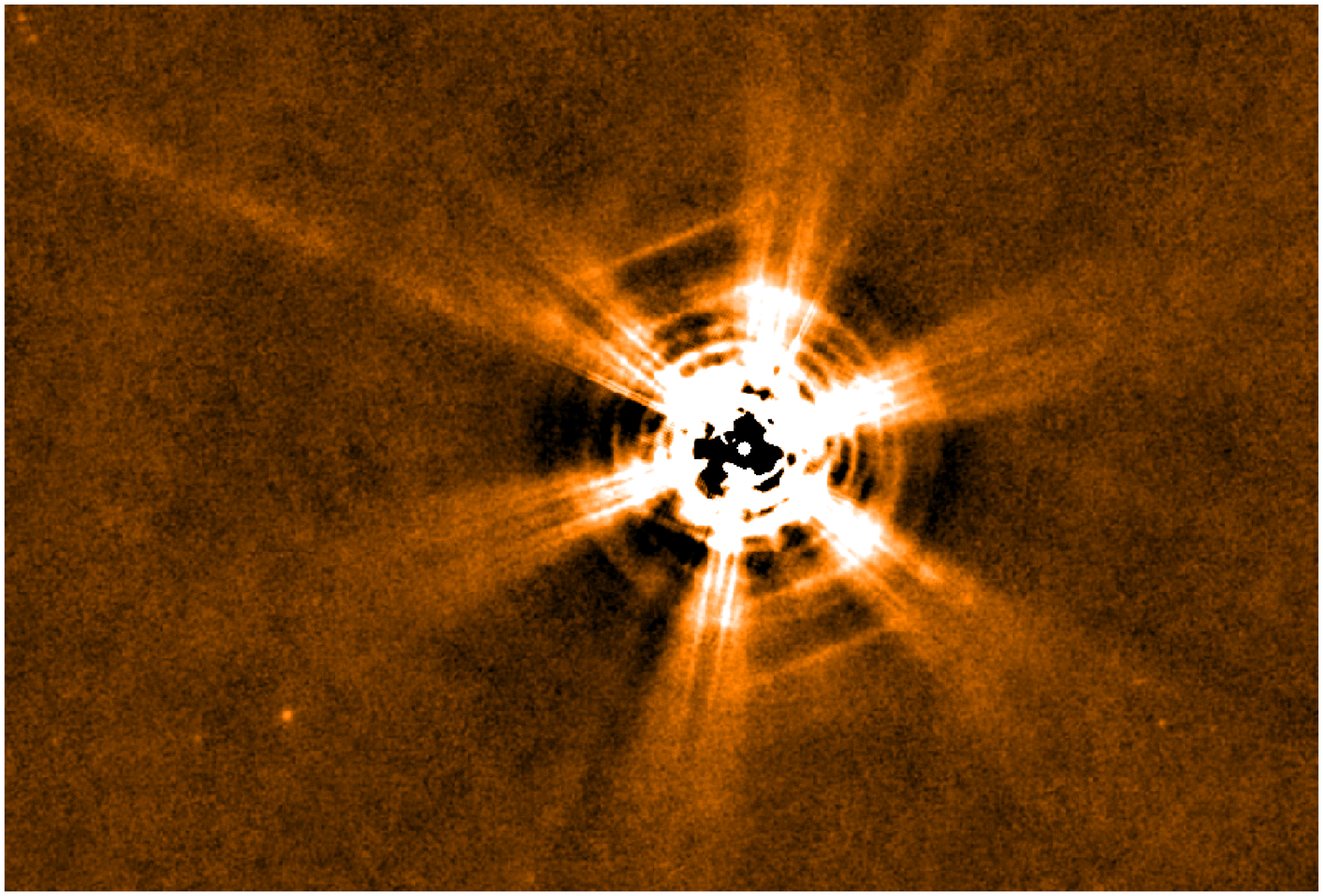}}}
    \end{minipage}
    \hfill
    \begin{minipage}[t]{.379\textwidth}
        \centerline{\resizebox{\textwidth}{!}{\includegraphics{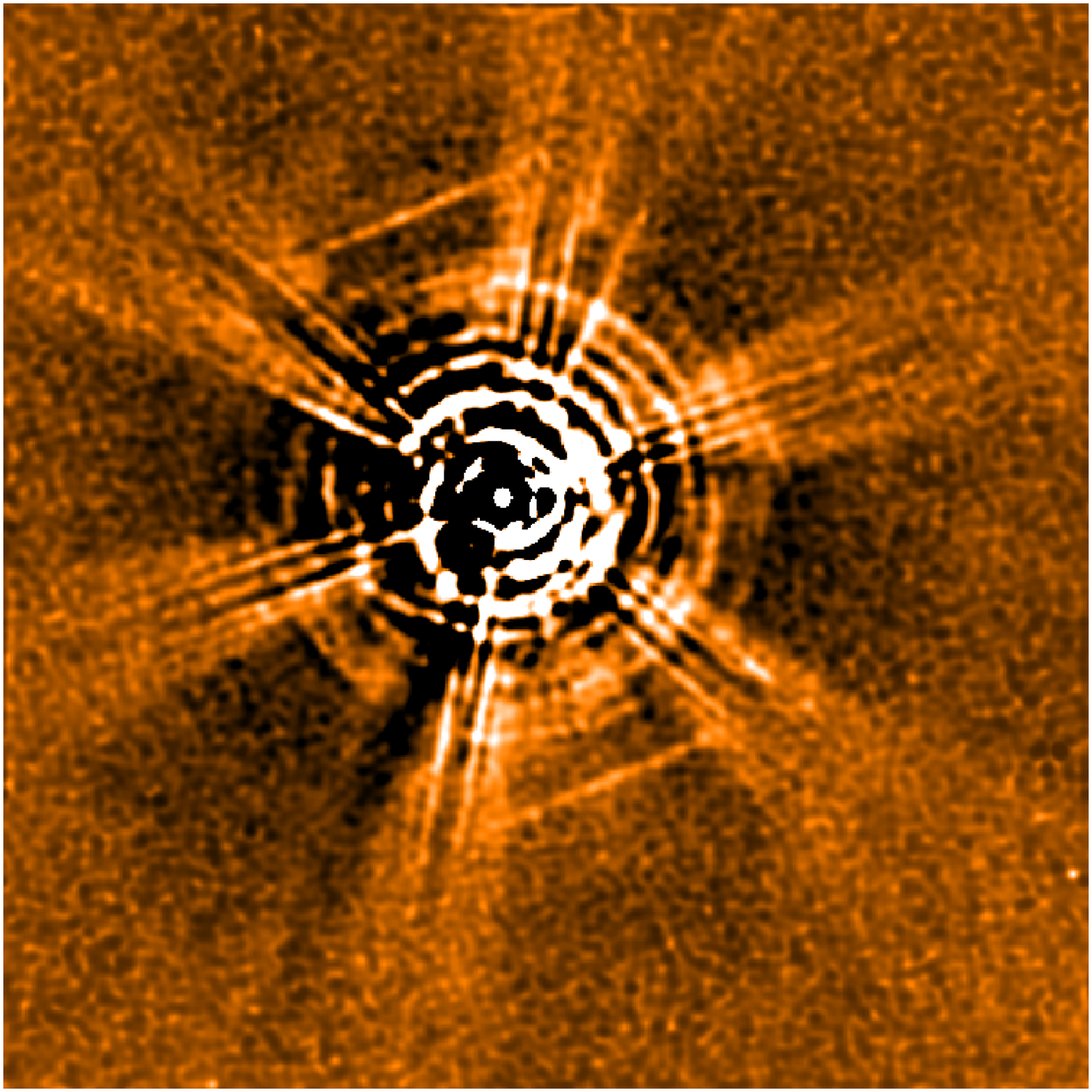}}}
    \end{minipage}

    \begin{minipage}[t]{.062\textwidth}
        \vspace*{-4cm}
        \centering{\textbf{\large{100\,$\mu$m}}}
    \end{minipage}
    \begin{minipage}[t]{.561\textwidth}
        \centerline{\resizebox{\textwidth}{!}{\includegraphics{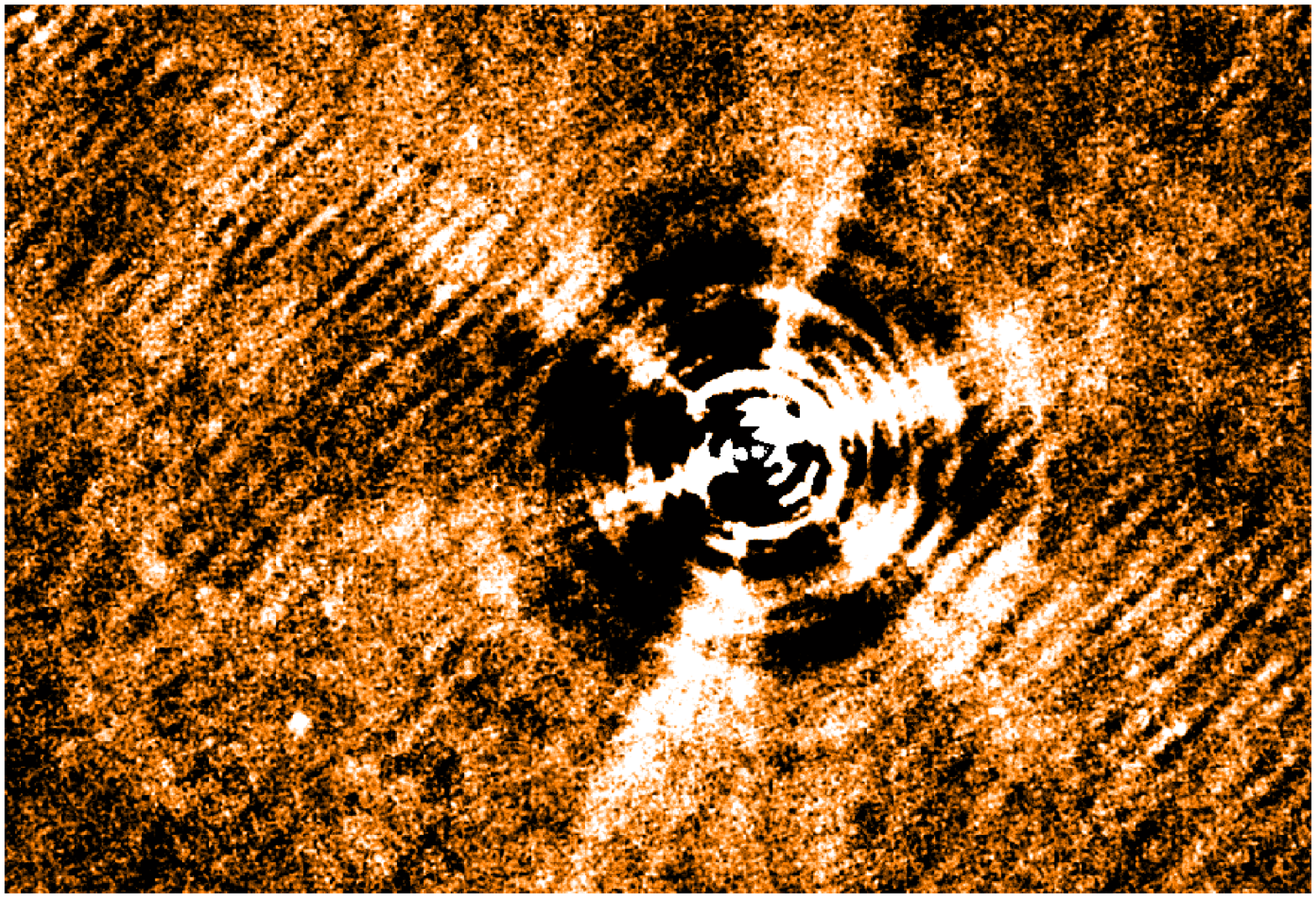}}}
    \end{minipage}
    \hfill
    \begin{minipage}[t]{.381\textwidth}
        \centerline{\resizebox{\textwidth}{!}{\includegraphics{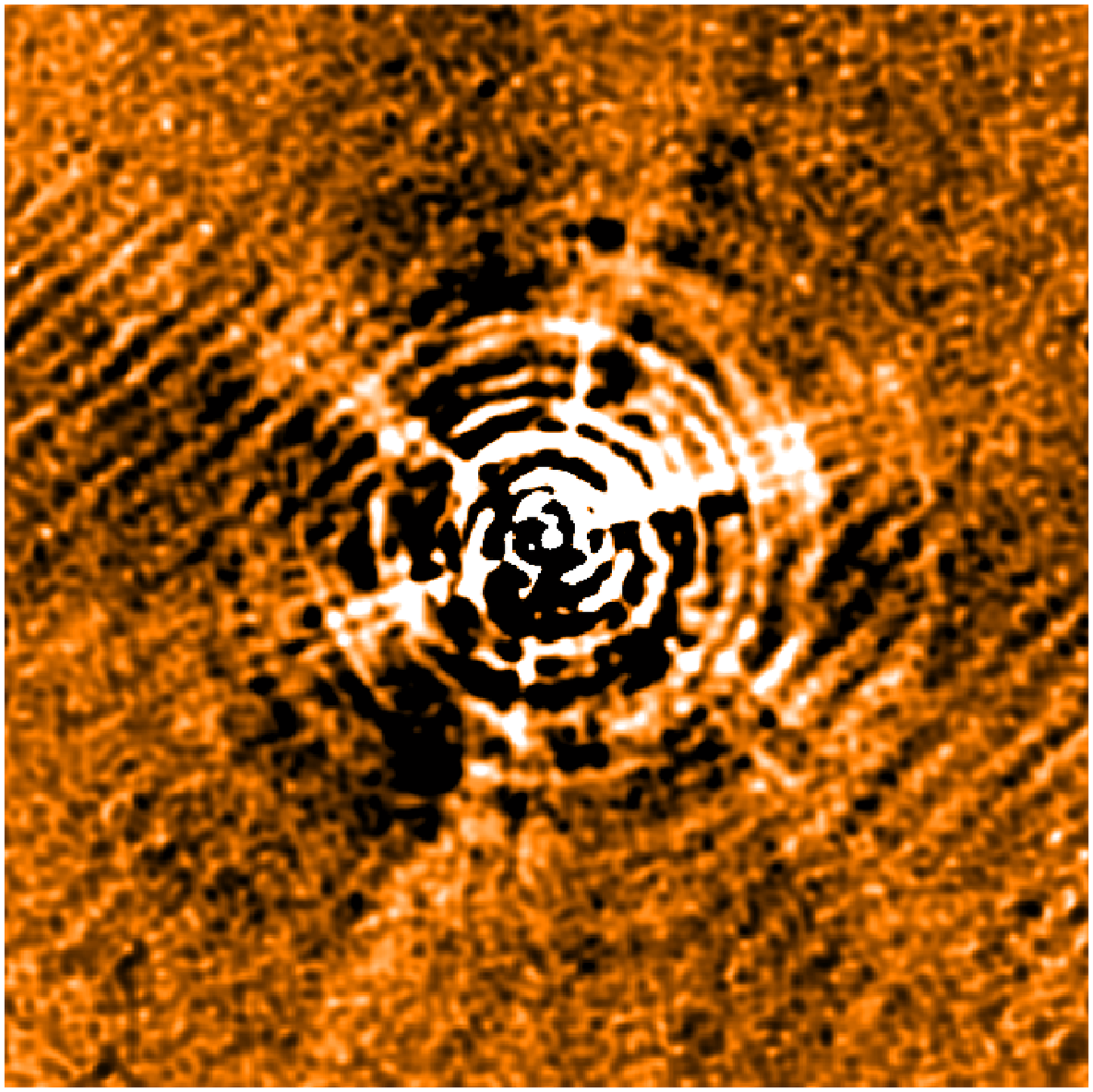}}}
    \end{minipage}
   
     \begin{minipage}[t]{.062\textwidth}
        \vspace*{-4cm}
        \centering{\textbf{\large{160\,$\mu$m}}}
    \end{minipage}
   \begin{minipage}[t]{.561\textwidth}
        \centerline{\resizebox{\textwidth}{!}{\includegraphics{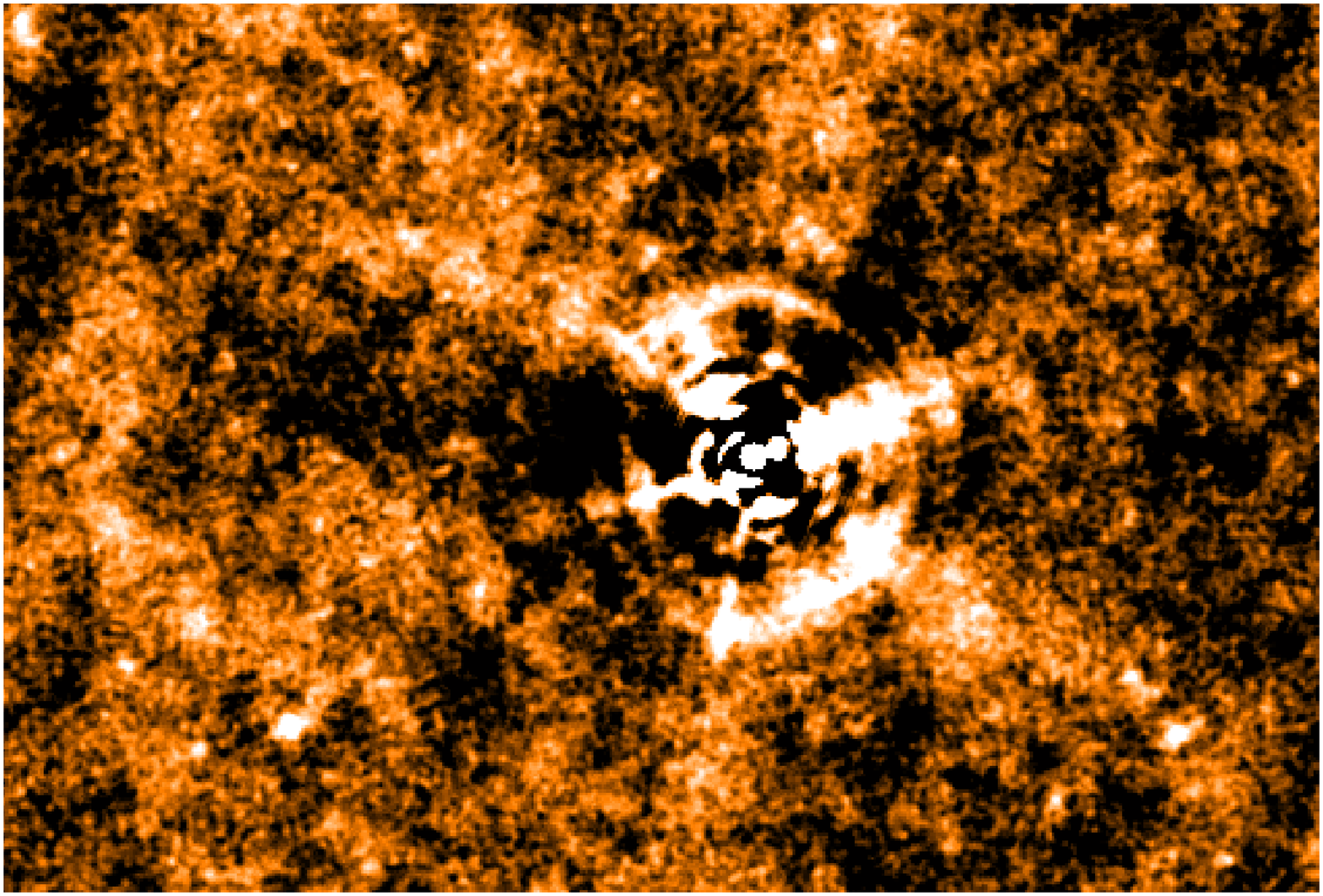}}}
    \end{minipage}
    \hfill
    \begin{minipage}[t]{.379\textwidth}
        \centerline{\resizebox{\textwidth}{!}{\includegraphics{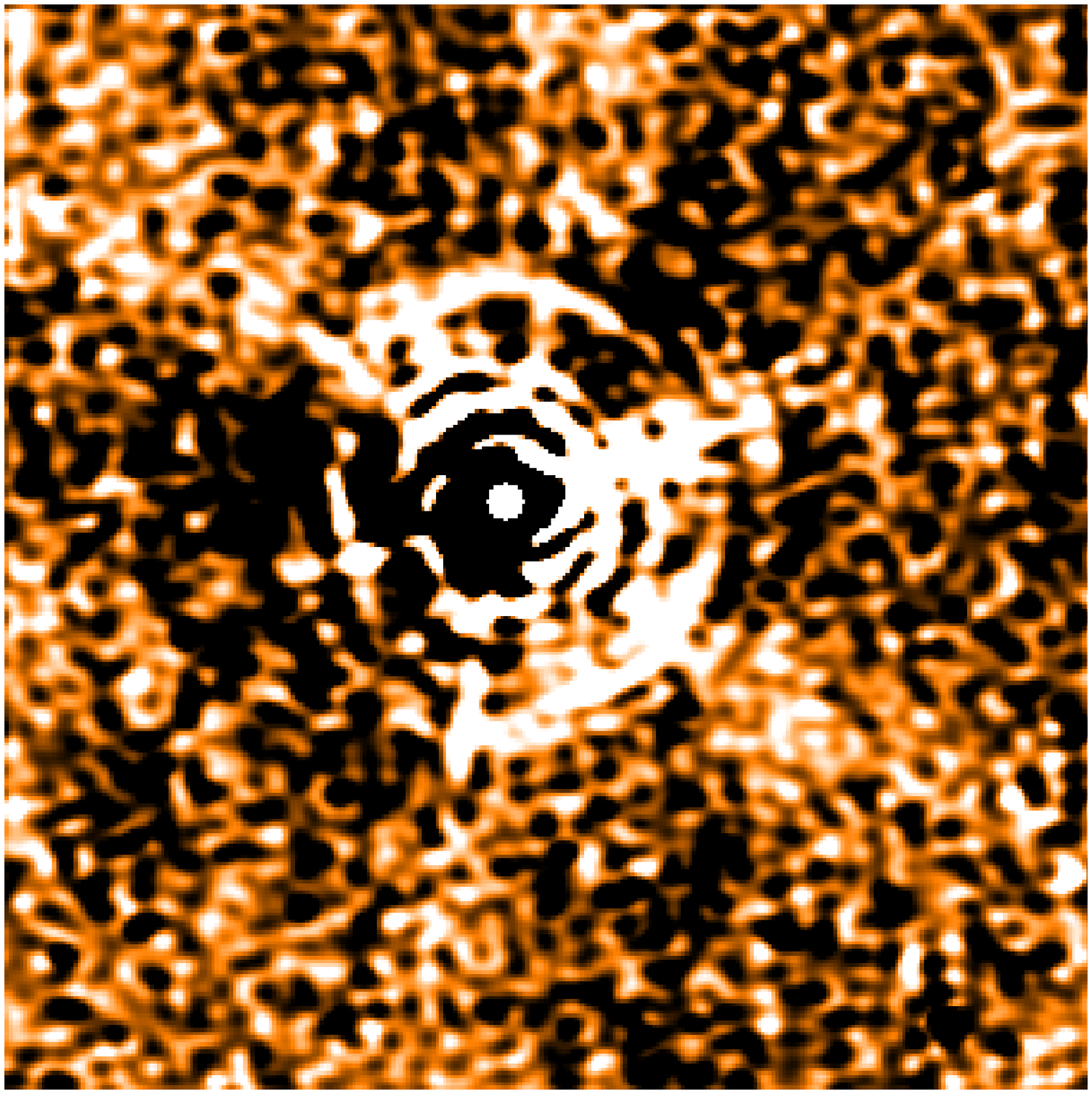}}}
    \end{minipage}
\caption{PACS images of IRC\,+10216 at 70\,$\mu$m (upper), 100\,$\mu$m (middle), and 160\,$\mu$m (bottom) after subtraction of the smooth halo of the circumstellar envelope. Left-hand column shows the data without deconvolution, right-hand column when deconvolution is applied before the subtraction of the smooth envelope structure.
The field-of-view (FOV) for the left-hand column is 16\arcmin x 11\arcmin, while for the deconvolved images the FOV is 10\arcmin x 10\arcmin. North is at the top, east is to the left. The pixel size is 1\arcsec\ for the 70 and 100\,$\mu$m images and is 2\arcsec\ for the 160\,$\mu$m image.  The six radial spikes are due to the support spider structure of the secondary mirror; these regions are not taken into account in our deeper analysis (see also Fig.~\ref{FIG_PACS100}).  The diagonal striping in the 100\,$\mu$m image is instrumental. The innermost arc is found at 24\arcsec, the outermost one at 320\arcsec. The bowshock detected at ~480\arcsec\ to the east by \citet{Ladjal2010A&A...518L.141L} can be seen in the left-hand image at 160\,$\mu$m.}
\label{FIG_PACS}
\end{figure*}

The observations were obtained using the Photodetector Array Camera and Spectrometer \citep[PACS,][]{Poglitsch2010A&A...518L...2P} on board the Herschel satellite \citep{Pilbratt2010A&A...518L...1P}, and are part of the MESS guaranteed time key programme \citep{Groenewegen2011A&A...526A.162G}, which is investigating the dust and gas chemistry of a large sample of post-main-sequence stars. The scan-map observing mode was used with a scan speed of 20\arcsec/s. The total sky coverage is 30\arcmin$\times$30\arcmin. Two scan maps were taken with a scanning angle of 90\deg\ between the two scans to achieve a homogeneous coverage. The PACS 160\,$\mu$m data were already presented by \citet{Ladjal2010A&A...518L.141L}, who discussed the detection of a bowshock at $\sim$585\arcsec\ when measured from the nearest focus of an ellipse fit to the shape of the bowshock, or at 480\arcsec\ when measured from the central star.

The data reduction was performed using the \textit{Herschel Interactive Processing Environment} (HIPE) \citep[see][for more details]{Groenewegen2011A&A...526A.162G}. To remove low-frequency noise, causing additive brightness drifts, the {\sc{scanamorphos}} routine \citep{Roussel2011} was applied. Deconvolution of the data was done using the method as described in \citet{Ottensamer2011}. 

To emphasize the shell morphology, we removed the central extended envelope halo by subtracting a smooth, azimuthally averaged profile represented by a power law $r^{-\alpha}$ for $r<500$\arcsec. The power $\alpha$ was determined directly from the local minima in the data for each image separately, and represents the azimuthally averaged (pseudo-) continuum intensity. The power index $\alpha$ varies between 2.07 and 2.50. The power index is different from that for a simplified $r^{-1}$ power law, obtained in the case of an optically thin, homogeneous, spherical envelope with a constant dust temperature or a $r^{-1.5}$ power law when the dust temperature is proportional to $r^{-0.5}$. As discussed in more detail in Sect.~\ref{enhancements}, the smooth underlying intensity represents both the continuum intensity (see red dotted line in Fig.~\ref{Fig:density}) and the additive contribution from each shell, resulting in a pseudo-continuum with a higher value for $\alpha$.  

While the PACS instrument offers a pixel size of 3.2\arcsec\ in the 70 and 100\,$\mu$m bands and of 6.4\arcsec\ in the 160\,$\mu$m band, the images  shown in this paper are oversampled by a factor of 3.2, resulting in a sampling of 1\arcsec\ and 2\arcsec\ per pixel, respectively. The PACS resolution at 70, 100, and 160\,$\mu$m is, respectively, $\sim$5.5\arcsec, $\sim$6.7\arcsec, and $\sim$11\arcsec. The final PACS images are shown in Fig.~\ref{FIG_PACS}.


\section{Results} \label{Results}

\begin{figure}[htp]
 \includegraphics[width=0.48\textwidth]{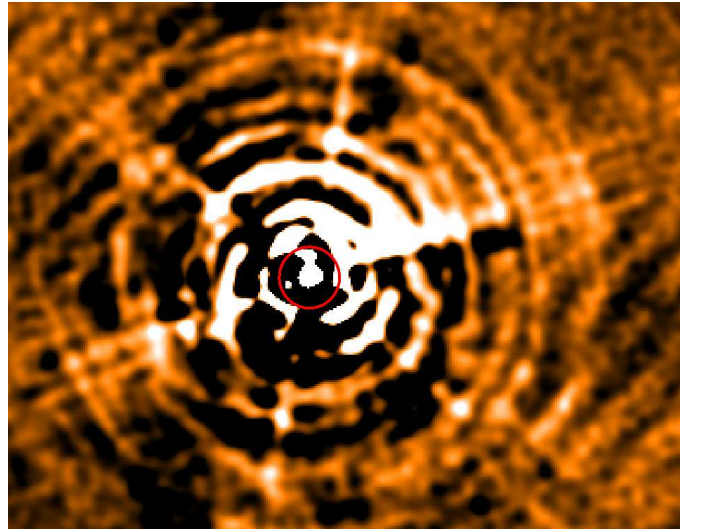}
\caption{Zoom into the inner 5\arcmin30\arcsec\ x 4\arcmin30\arcsec\ of the PACS 100\,$\mu$m image (cf.\ the middle panel in the right-hand column of Fig.~\ref{FIG_PACS}). The red circle with a radius of 15\arcsec\ marks the region where some PSF artefacts are still visible.}
\label{PSF}
\end{figure}

\begin{figure*}[htp]
\includegraphics[height=0.49\textwidth]{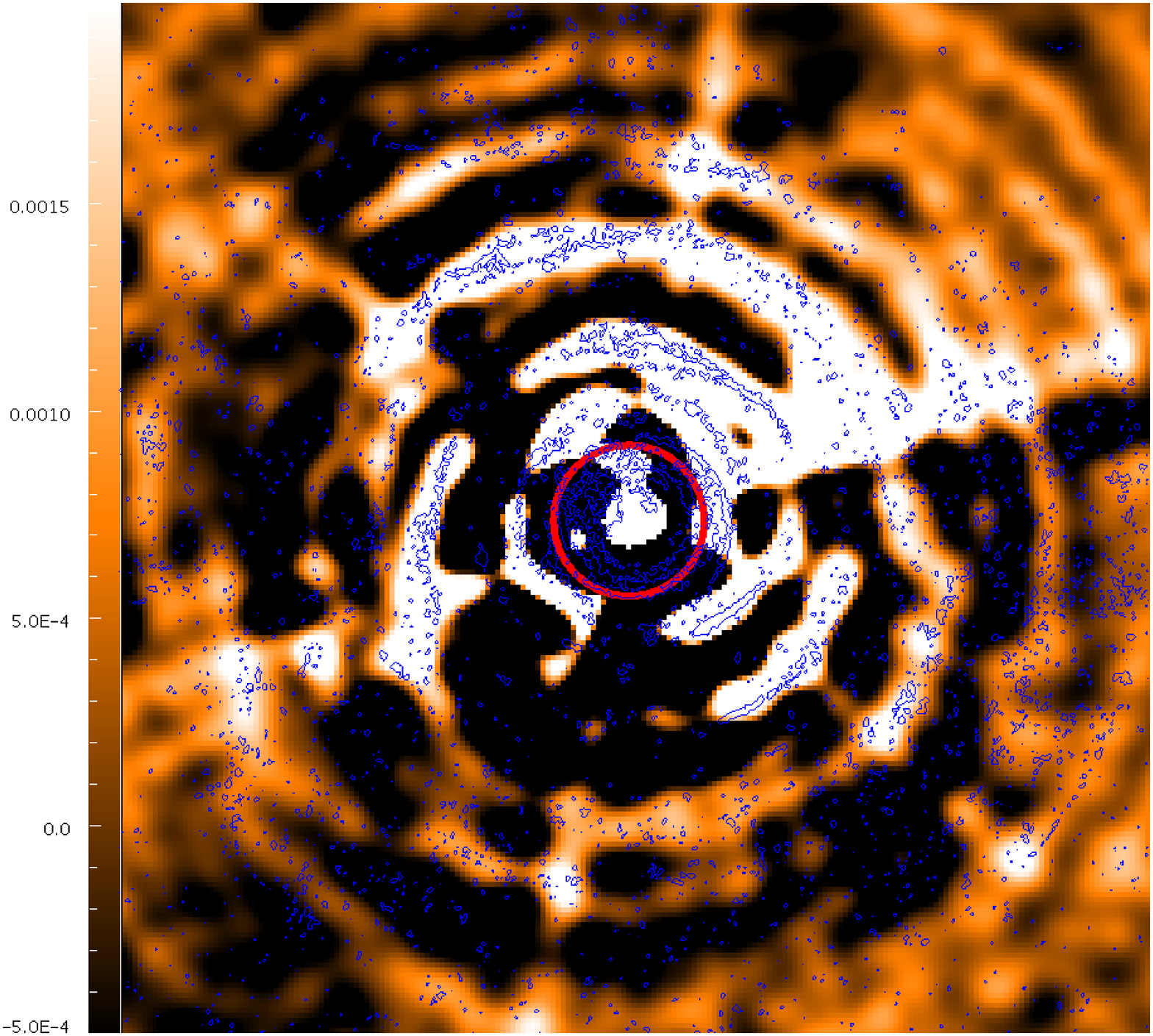}
\includegraphics[height=0.49\textwidth]{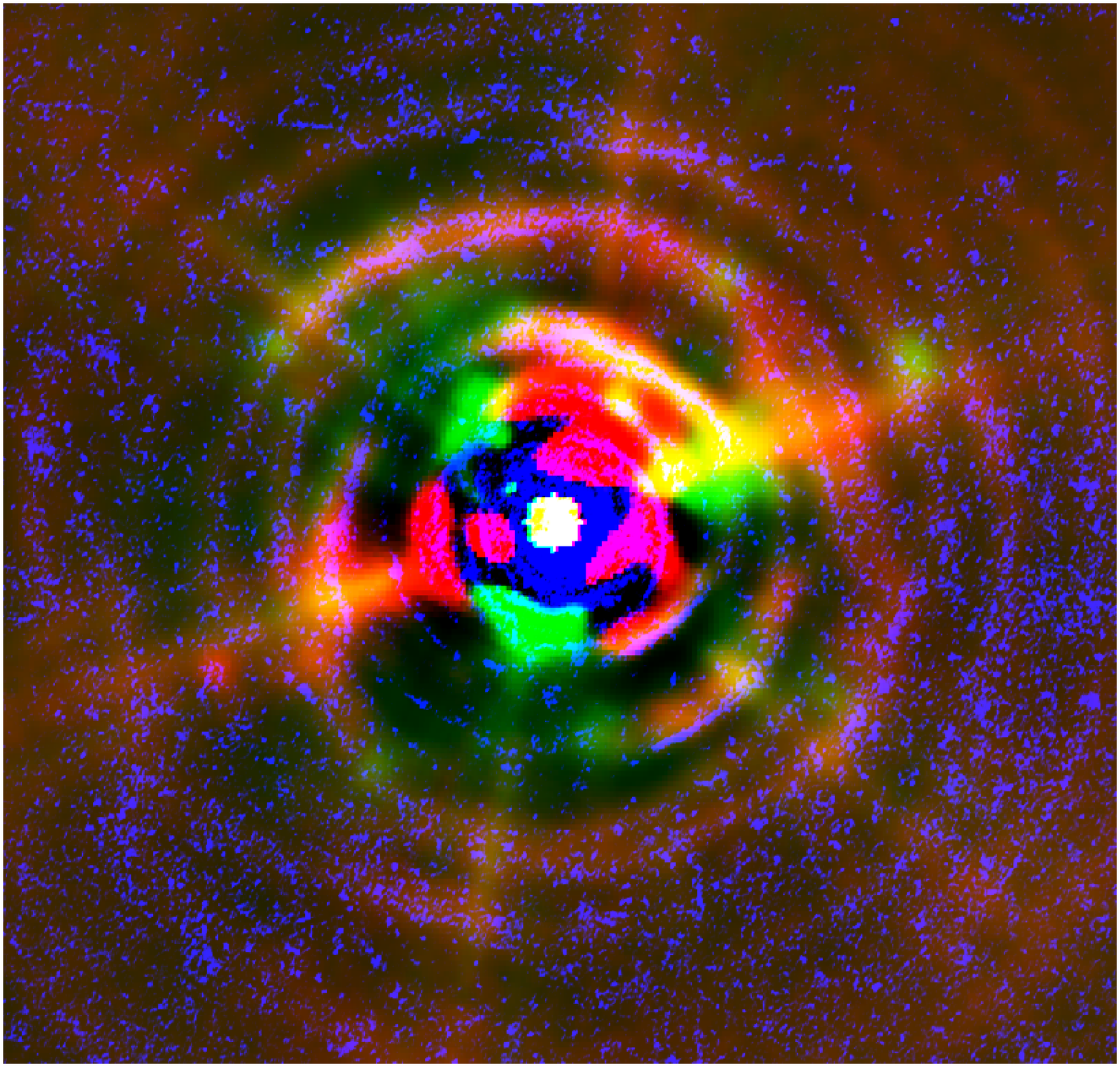}
\caption{Comparison between PACS images and deconvolved VLT/FORS1 V-band image \citep[see Fig.~4 in][]{Leao2006A&A...455..187L}. The field-of-view is 204\arcsec x 204\arcsec. \textit{Left-hand image:} Deconvolved PACS 100\,$\mu$m image (flux units being mJy/arcsec$^2$) with superposed contours (in blue) of the V-band image by \citet{Leao2006A&A...455..187L}. The red circle marks the region in which the flux is unreliable because of PSF artefacts (see Fig.~\ref{PSF}).
\textit{Right-hand image:}
Combined image of the PACS 70\,$\mu$m (green), PACS 100\,$\mu$m (red), and V-band image (blue). The outermost arcs detected in the V-band image (at a radius of $\sim$32\arcsec, 53\arcsec, and 73\arcsec) are clearly seen in the PACS images as well. (For the best result, viewing this figure electronically is recommended).}
\label{FIG_ALL}
\end{figure*}

 It is clear that the extended envelope of IRC+10216 shown in Fig.~\ref{FIG_PACS} is composed of multiple shells (or arcs). Thanks to the high sensitivity of PACS, at least a dozen shells are detected for the first time beyond a radius of 1~arcmin, with the outermost shell being located at $\sim$320\arcsec\ from the central star. For a terminal gas velocity, $v_\infty$, of 14.5\,km/s \citep{Decin2010A&A...518L.143D}, a distance of 150\,pc \citep{Crosas1997ApJ...483..913C}, and neglecting the drift velocity (see Sect.~\ref{origin}), the outermost arc was ejected about 16\,000\,yr ago. The shell separation is typically between $\sim$10\arcsec\ and 35\arcsec\ (or 500\,--1\,700\,yr), but some shells intersect each other. Owing to the complex point-spread function (PSF) and the bright contribution from the central target at these infrared wavelengths, the innermost shell detected by PACS is (only) at $\sim$24\arcsec (see Fig.~\ref{PSF}), while \citet{Leao2006A&A...455..187L} detected the first shells as close as 4\arcsec\ from the central target using V-band images. A good correspondence is seen between the innermost arcs seen in the PACS images and the  outermost shells seen in the V-band data of \citet{Leao2006A&A...455..187L} (see Fig.~\ref{FIG_ALL}). As was the case for the inner shells \citep[$<$1\arcmin, e.g.][]{Mauron1999A&A...349..203M,Leao2006A&A...455..187L}, all shells in the PACS images appear to be non-concentric and azimuthally incomplete. Some arcs have a positive and others a negative inclination. However, while the arcs seen in the V-band image cover $\sim$30\deg--90\deg\ in azimuth, the arcs in the PACS 100\,$\mu$m image extend $\sim$45\deg--200\deg. The source IRC\,+10216 was also observed with the SPIRE instrument in the framework of the MESS GTKP. However, after applying the same reduction strategy, only two faint arcs at $\sim$1\arcmin\ and 2\arcmin\ are visible in the SPIRE 250\,$\mu$m image (see Fig.~\ref{FIG_SPIRE250} in the Appendix).

 The arcs seen in the B and V band images presented by \citet{Mauron1999A&A...349..203M} and \citet{Leao2006A&A...455..187L} are caused by dust scattering of the external illumination by the ambient Galactic light, but the PACS images show the direct emission of dust grains of which the main heating source is the thermal emission of nearby dust grains, with minor contributions from either the stellar UV flux (close to the star) or the diffuse Galactic light (in the outer envelope). The highest contrast between the arcs is reached in the PACS 100\,$\mu$m image. Applying aperture photometry on the azimuthally averaged deconvolved PACS images and using a  modified blackbody of the form $B_\nu \cdot \lambda^{-\beta}$, as expected for a grain emissivity $Q_{\rm{abs}} \sim \lambda^{-\beta}$ with $\beta$ equal to 1.2 \citep[representing amorphous carbon,][]{Mennella1998ApJ...496.1058M}, we derived a dust temperature between 108$\pm$5\,K at 20\arcsec\ and 40$\pm$5\,K at 180\arcsec. The temperature of the bowshock at 480\arcsec\ is $25\pm3$\,K \citep{Ladjal2010A&A...518L.141L}.

The thickness of the shells  typically varies within the range $\sim$5\arcsec--8\arcsec, and most of them are hence not resolved. Local variations in one arc are seen.  The typical thickness derived from the V-band image by \citet{Leao2006A&A...455..187L} is around 1.6\arcsec$\pm$0.3\arcsec (obtained for a pixel size of 0.2\arcsec\ and a spatial resolution of $\sim$0.6\arcsec), and no relation was found between the shell thickness and the distance to the center, which could be expected from the natural expansion of the envelope.


\citet{Dinh2008ApJ...678..303D} found that the gas shells seem to be more concentrated on the western part of the envelope and scarce at position angles of $\sim$0\deg--30\deg\ (measured counterclockwise from north). The PACS observations confirm the higher concentration of shells in the western part, but the lowest number of arcs close to the target are found in the eastern part (and not in the north-eastern region; see also Fig.~\ref{FIG_ENHANCEMENT}).


\vspace*{-.4cm}
\section{Discussion} \label{Discussion}

 \subsection{Shell intensity and density contrast of the 100\,$\mu$m image} \label{enhancements}

\begin{figure}[htp]
        \centerline{\resizebox{0.48\textwidth}{!}{\includegraphics{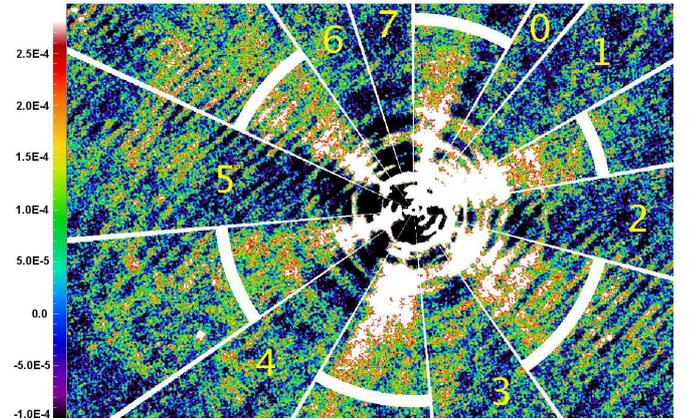}}}
        \caption{PACS 100\,$\mu$m image (in units of mJy/arcsec$^2$) of IRC\,+10216 after subtraction of the smooth halo of the circumstellar envelope. The field-of-view is 975\arcsec x 660\arcsec. North is at the top, east is to the left. The pixel size is 1\arcsec. The central target  is marked by the innermost white circle. The six radial spikes indicated by the white sectors are due to the support spider structure of the secondary mirror; these regions are not taken into account in the analysis. The position angles of each sector used in the analysis are listed in Table~\ref{sectors}.}
        \label{FIG_PACS100}
\end{figure}

\begin{figure}[htp]
 \includegraphics[width=0.48\textwidth]{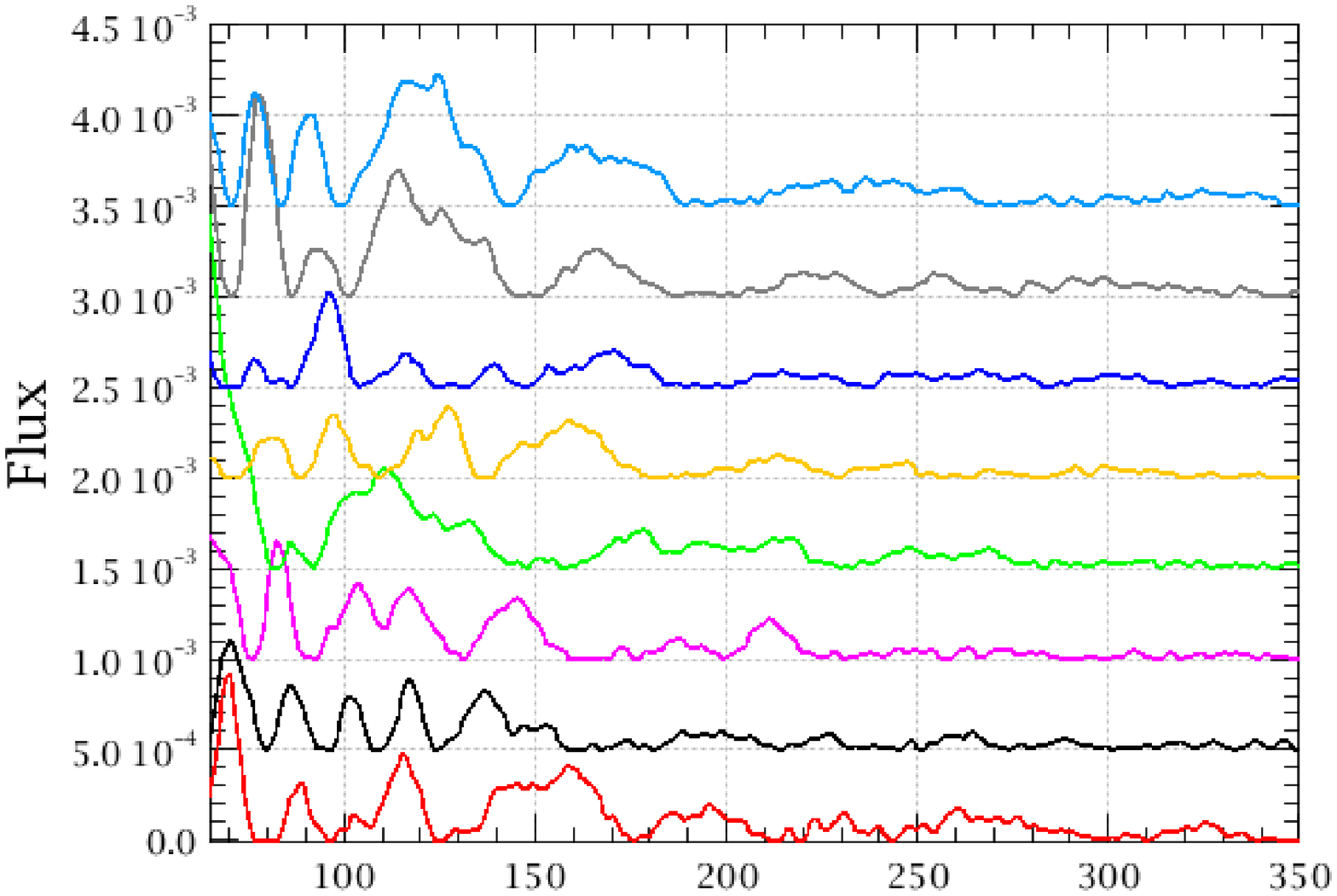}
 \includegraphics[width=0.48\textwidth]{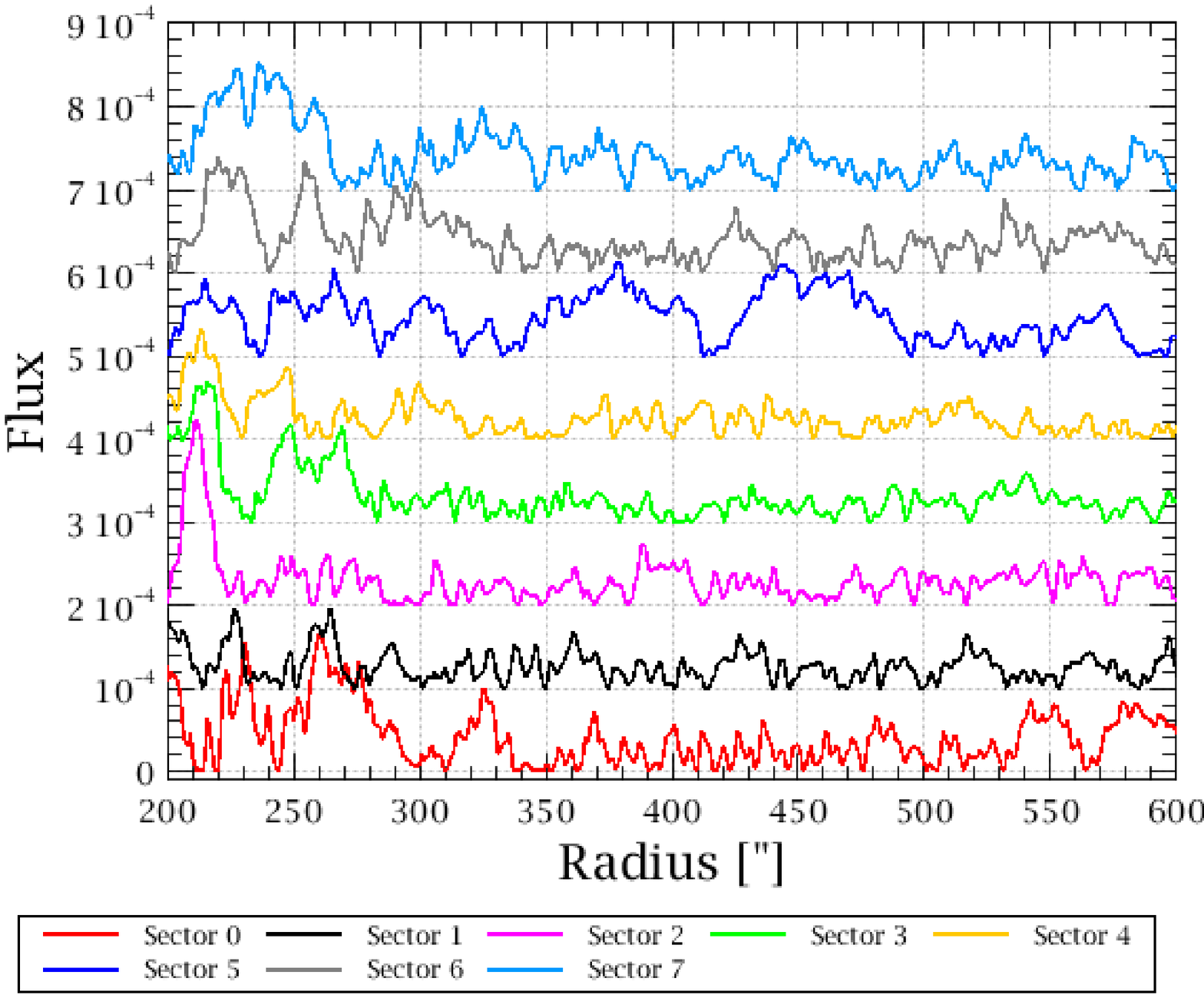}
\caption{Mean intensity enhancements (in Jy/arcsec$^2$) in each sector of the PACS 100\,$\mu$m image (see Fig.~\ref{FIG_PACS100}). The position angle covered by each sector is listed in Table~\ref{sectors}. The different sectors are offset to each other. The upper panel shows the intensity (in Jy/arcsec$^2$) in the arcs between 65\arcsec\ and 350\arcsec\ away from the central target, while the lower panel shows the outermost arcs until 600\arcsec\ away. The bowshock as discussed by \citet{Ladjal2010A&A...518L.141L} is seen at $\sim$450\arcsec\ in sector 5 (see also Sect.~\ref{sec_bowshock} in the Appendix). }
\label{FIG_ENHANCEMENT}
\end{figure}

From the PACS images, the relative intensity variations in the arcs can be derived. This is illustrated in the case of the 100\,$\mu$m image where the arcs  are most clearly visible.  We only used regions unaffected by PSF features caused by the support spider structure of the secondary mirror (see Fig.~\ref{FIG_PACS100} and Table~\ref{sectors}). To measure the intensity variations along the radial intensity profiles between the different local minima and maxima, the PACS images were projected from polar to cartesian coordinates \citep[see, e.g., Fig.~6 in ][]{Leao2006A&A...455..187L}.
The mean intensity enhancements in the different sectors  are displayed in Fig.~\ref{FIG_ENHANCEMENT}.
The ratio of the integrated intensity to the full envelope intensity in the individual shells varies between $\sim$2 and 15\%, with the brightest arcs seen in sector 0. The lowest numbers of arcs is seen in sector 5.
 While the typical thickness of the shells  varies within the range $\sim$5\arcsec--8\arcsec, the shells appear broader in Fig.~\ref{FIG_ENHANCEMENT} because of the non-concentric nature of the shells.

\begin{table}[htp]
\caption{Position angles (PA; measured counterclockwise from north) of the different sectors used in the analysis of the intensity variations in the arcs (see Fig.~\ref{FIG_PACS100}). }
\label{sectors}
 \begin{tabular}{lc|lc}
  \hline \hline
Name & PA & Name & PA \\
\hline
Sector 0 & 320\deg--330\deg & Sector 4 & 125\deg--150\deg \\
Sector 1 & 300\deg--330\deg & Sector 5 & \ 65\deg--95\deg \\
Sector 2 & 255\deg--280\deg & Sector 6 & 17.5\deg--35\deg \\
Sector 3 & 185\deg--215\deg & Sector 7 & \ \ 0\deg--17.5\deg  \\
\hline
 \end{tabular}
\end{table}

\begin{figure}[htp]
\includegraphics[width=0.48\textwidth]{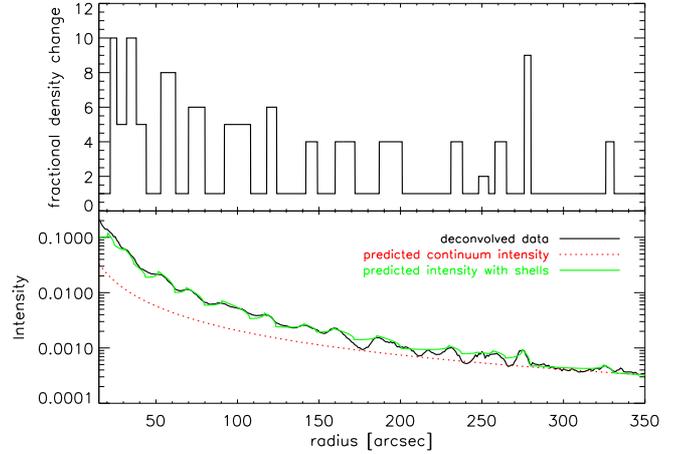}
\caption{Multiple-shell radiative-transfer model fit to the radial intensity profile at a PA of 325\deg\ in the 100\,$\mu$m image. \textit{Bottom panel:} The black line shows the deconvolved observations, the red dotted line shows the predicted continuum intensity based on Eq.~(\ref{Eq1}), and the green full line shows the multiple shell fit to the data based on Eq.~(\ref{Eq1}) using the fractional density variations shown in the upper panel. \textit{Upper panel:} Relative density change $f(r)$ in each arc. The profile was not fitted for $r<15$\arcsec, because of the complex PSF structure in this region.}
\label{Fig:density}
\end{figure}

As shown by \citet{Mauron2000A&A...359..707M}, the shell density contrast can be obtained from the shell intensity profiles. To illustrate this, we modelled the extended radial intensity profile taken at a position angle (PA) of 325\deg\ in the 100\,$\mu$m image (see Fig.~\ref{Fig:density}). The smooth envelope structure is modelled assuming a constant mass-loss rate, i.e.\ the density varies as $r^{-2}$ (or $\rho(r)=a \cdot r^{-2}$, with $a=\dot{M}/(4 \pi v_\infty)$). The shell-intershell contrast is characterized by $\rho(r) = f(r) \cdot a \cdot r^{-2}$, where $f(r)$ represents the relative change in density in each region. As discussed by \citet{Mauron2000A&A...359..707M}, the modelling assumption that the shells are complete, might overestimate their contribution to the radial profile since the arcs tend to be prominent over a limited azimuthal angle. This is, however, roughly compensated by the additional contribution of the other shells which are not seen as prominent arcs when they do not intersect the plane of the sky. The thermal emission of the dust grains is modelled assuming {a spherically symmetric envelope and} LTE (local thermodynamic equilibrium), i.e.\ the intensity $I_\nu$ can be written as
\begin{equation}
 I_\nu = I_{\rm bg} e^{-\tau_\nu} + S_\nu (1-e^{-\tau_\nu})\,,
\label{Eq1}
\end{equation}
with $I_{\rm{bg}}$ being the interstellar radiation field taken from \citet{Porter2006ApJ...648L..29P}, $\tau_\nu$ the optical depth along the line of sight, and $S_\nu$ the source function assumed to be characterized by $B_\nu(T_{\rm{dust}}(r))$. {The fractional contribution of the first term on the right-hand side of Eq.~(\ref{Eq1})  to the second term is $<$0.4\% for $r<350$\arcsec, and can therefore be neglected.} The dust temperature distribution is taken from the SED modelling presented in \citet{Decin2010Natur.467...64D}{, and can be approximated as
\begin{equation}
 T_{\rm dust}(r) = 848 * \left( r/r_{\rm inner} \right)^{-0.48}\,,
\label{Eq_Tdust}
\end{equation}
with $r_{\rm inner}$ the dust condensation radius at 0.127\arcsec. This smooth (analytical) dust temperature distribution yields values lower than the dust temperatures derived in the shells from the aperture photometry in Sect.~3. This is because the functional form of a modified blackbody is only valid in the case of optically thin emission assuming a constant dust temperature along the line-of-sight. The more sophisticated dust modelling presented in \citet{Decin2010Natur.467...64D} takes optical depth effects into account when considering a radially varying dust temperature. }

The radial intensity profile shown in the bottom panel of Fig.~\ref{Fig:density} was fitted using the fractional density variation displayed in the upper panel of Fig.~\ref{Fig:density}. Density contrasts of up to a factor of ten are seen in some arcs, but a more typical value is about four. This is in good agreement with the results deduced from the inner arcs in the scattered V-band image by \citet{Mauron2000A&A...359..707M}. Our results indicate that the arcs contain $\sim$50\% more dust mass than the smooth envelope.

\subsection{Tracing the mass-loss history of IRC+10216}\label{Mdothistory}
\citet{Sahai2010ApJ...711L..53S}, \citet{Ladjal2010A&A...518L.141L}, and \citet{Cox2011} discussed the detection of a bowshock eastward of the central star, which is the result of the interaction between the AGB wind and the interstellar medium (ISM). From the ultraviolet (UV) images, \citet{Sahai2010ApJ...711L..53S} deduced the approximate locations of the termination shock\footnote{where the stellar wind velocity  goes from supersonic to subsonic values}, astropause\footnote{where the ram pressure of the ISM equals the ram pressure of the stellar wind}, and bowshock\footnote{where the ISM velocity goes from supersonic to subsonic values} eastward of the central target, to be at $\sim$500\arcsec, 590\arcsec\ and 670\arcsec, respectively. 


The PACS images show a clear intensity enhancement around 450\arcsec\ at 100\,$\mu$m (see Fig.~\ref{FIG_ENHANCEMENT}) and around 480\arcsec\ at 160\,$\mu$m (see Fig.~\ref{FIG_BOWSHOCK} in the Appendix). As shown by \citet{Ladjal2010A&A...518L.141L}, a dust temperature of $25\pm3$\,K is derived from the PACS and SPIRE images for a mixture of carbon-rich (from the stellar wind) and oxygen-rich (from the ISM) material ($\beta$=1.6). This suggests that the PACS images detect the turbulent astropause, where instabilities mix the ISM and circumstellar material and that the termination shock is situated $<$450\arcsec.  
Hydrodynamical simulations with the AMRVAC-code \citep{vanMarle2011} show that the termination shock is indeed closer to the star, i.e.\ at 390\arcsec\ and that the bowshock  is located at $\sim$650\arcsec\ (in accordance with the observations). The region interior to the termination shock consists of the unshocked, freely streaming stellar wind. In the turbulent astropause, Rayleigh-Taylor and Kelvin-Helmholtz instabilities dominate the picture.

Although the very faint emission makes it impossible to detect dust shells beyond 320\arcsec\ in the PACS images, it is reasonable to assume that the nested shell structure extends all the way out to the bowshock (see  Fig.~\ref{FIG_ENHANCEMENT}). The imprint of this complex mass-loss history will be traceable until the termination shock (at 390\arcsec\ or during a period of 19000\,yr). This part of the envelope has a mass of  $\sim$0.19\,\Msun\ for a mean mass-loss rate of $1 \times 10^{-5}$\,\Msun/yr. However, in the turbulent astropause region, any density contrast between the shell and intershell regions will be wiped out by the instabilities. For a mean velocity of 1\,km/s  in the region between termination shock and bowshock (as deduced from the hydrodynamical simulations), this means that at least the first $\sim$200\,000\,yr of mass-loss history are lost. After undergoing mass loss for at least 219\,000\,yr, the total CSE mass is estimated to be higher than 2\,\Msun, which is in accordance with the initial mass of 3\,$\la$\,M\,$\la$\,5\,\Msun\ as estimated by \citet{Guelin1995A&A...297..183G}.

Using aperture photometry, one can also estimate the total dust (and gas) mass in the envelope, and compare that to the value of  0.19\,\Msun\  derived in the previous paragraph. The total --- background-subtracted --- flux in an aperture of radius 390\arcsec\ is $\sim$1750\,Jy, of which some 30\% comes from the radius $<$15\arcsec. A first order estimate of the dust mass is 
\begin{equation}
M_{\rm dust} = \frac{ F_\nu \, \lambda^2 \, d^2 }{2\, k \,T_{\rm dust}\,
\kappa_\lambda}\,,
\label{Eq2}
\end{equation}
\citep{Li2005AIPC..761..123L} with $F_\nu$ the observed flux (in erg s$^{-1}$ cm$^{-2}$ Hz$^{-1}$), $\lambda$ the
wavelength (in cm), $d$ the distance (cm), $k$ the Boltzman constant (= 1.381 10$^{-16}$\,erg K$^{-1}$),
$T_{\rm dust}$  the dust temperature (in K) taken to be 30\,K (see Sect.~\ref{Results}), and $\kappa_\lambda$ the dust opacity at the wavelength observed. We adopted a dust absorptivity at 100\,$\mu$m of 25\,cm$^2$ g$^{-1}$ \citep{Hildebrand1983QJRAS..24..267H}, but other higher values are also found in the literature \citep[e.g.,][]{Mennella1998ApJ...496.1058M}. Using Eq.~(\ref{Eq2}), the total estimated dust mass for the envelope within 390\arcsec\ is $M_{\rm dust}  \sim\ 9 \times 10^{-4}$\,M$_\odot$. Adopting a gas-to-dust ratio of 250 \citep{Decin2010Natur.467...64D}, the total (gas+dust) envelope mass (for $r<$390\arcsec) is $\sim$0.23\,M$_\odot$. Although in good agreement with the value of 0.19\,\Msun\ derived in previous paragraph, one should realize that the uncertainty in this value is at least a factor of two owing to \textit{(1.)} the complex PSF yielding an uncertain flux estimate for $r<$15\arcsec, and \textit{(2.)} the uncertainties in dust opacities that depend strongly on the shape and the size of the grains.

\subsection{Origin of the shells}\label{origin}

We note that similar circular shells are seen around the carbon-rich post-AGB object AFGL~2688 \citep{Sahai1998ApJ...493..301S}. As noted already by other authors on the topic of the shell structure in IRC+10216, it is difficult to identify a physical mechanism that can explain the origin of the non-concentric incomplete shells, which occur at time separations between a few hundred yr and $\sim$2000\,yr. This time interval is much longer than the normal stellar pulsation period of 649 days \citep{LeBertre1992A&AS...94..377L}, and at the same time much shorter than the interpulse period between helium flashes, which for IRC+10216 is estimated to be between 6\,000 and 33\,000\,yr \citep{Ladjal2010A&A...518L.141L}.  

Three different types of models have been proposed in the literature. One type is based on the effects of a binary companion. However, as discussed in detail by \citet{Mauron2000A&A...359..707M}, the irregular arc spacing seen in the optical and infrared images does not support this type of model.

A different scenario is proposed by \citet{Soker2000ApJ...540..436S}, who suggested that a solar-like magnetic activity cycle is behind the formation of the multiple arcs. Owing to the appearance of magnetic cool spots during the active phase, the gas temperature can be locally  reduced. This can enhance dust formation and thus lead to a higher mass-loss rate from magnetic cool spots. The magnetic cool spots cover only a fraction of the stellar surface, and this inhomogeneity might explain the clumpiness seen on smaller spatial scales and in the arcs. Starspots have not yet been observed on AGB stars, although magnetic field strengths around only a few late-type stars have been measured \citep[e.g.,][]{Vlemmings2005A&A...434.1029V}.  \citet{Soker1999MNRAS.307..993S} suggest that cool magnetic spots on the surface of AGB stars might have a radius of $<$0.1R$_\star$. This would imply that several large starspots should co-exist at the same time sufficiently close to each other to explain the result obtained by \citet{Dinh2008ApJ...678..303D} that each shell typically covers 10\% of the stellar surface at the time of ejection. However, taking dynamical effects into account, this starspot model  would lead to a periodicity of the shells of half of the magnetic cycle \citep{Garcia2001ApJ...560..928G}. This periodicity contradicts the irregular shell spacings observed. 

\citet{Simis2001A&A...371..205S} and \citet{Sandin2004A&A...413..789S} studied the effect of drift-dependent dust formation. An intricate nonlinear interplay between gas-grain drift, grain nucleation, radiation pressure, and envelope hydrodynamics can result in gas and dust density shells that occur at irregular periods of a few hundred years. As explained by \citet{Simis2001A&A...371..205S}, the shells in their models appear perfectly round owing to the assumption of spherical symmetry. The fact that the chemical-dynamical system regulating the behaviour of the envelope is extremely stiff and reacts violently to all kind of changes suggests that in nature the shells will be incomplete or clumpy. However,   the predicted shell density variations are  much larger than the observed ones. This might be because clumpiness is not accounted for in this spherically symmetric models (see also next paragraph).

Some constraints are set by the (observational) link between the inner regions, which for both IRC\,+10216 and AFGL~2688 display a peanut- or bipolar-like structure in a clumpy environment and the spherical-like shells observed on larger scales. According to \citet{Menut2007MNRAS.376L...6M}, the change from roughness on small scales to smoothness on larger scales can be explained by  the dilution of the dust clumps while they move outward, or a shadowing effect in combination with multi-scattering of the stellar and dust-emitted photons. In addition, backwarming effects during non-isotropic dust formation will inhibit the formation of new dust clumps between the star and already formed dust clumps. However, when a dust clump has moved outward or in any other direction where the number of nearby dust clumps is lower, dust formation is still possible. Repeated over a longer period, this process might result in spherical-like shells.  This idea is supported by the high angular resolution observations of the HC$_3$N and HC$_5$N molecules, which have a clumpy and quasi-spherical distribution \citep{Dinh2008ApJ...678..303D}. The location of the arcs of the cyanopolyyne molecules coincide very closely with the dust arcs identified by \citet{Mauron1999A&A...349..203M} and later on by \citet{Leao2006A&A...455..187L}, suggesting that the drift velocity is zero during these phases of higher dust and gas density. From the observations, \citet{Dinh2008ApJ...678..303D} concluded that the arcs are true three-dimensional expanding features, and not projections on the plane of the sky of just a few larger three-dimensional shells as suggested by \citet{Fong2003ApJ...582L..39F}. Moreover, their velocity channel maps suggest that there might be slight variations in either the ejection velocity or ejection time between different parts of the shell, which might explain the non-circular nature of the arcs.

The PACS observations in combination with already published optical, infrared, and sub-millimetre images of the envelope around IRC+10216 suggest again that the dust formation and the onset of the mass loss is highly anisotropic, and that local density enhancements prevail until the termination shock. Local thermodynamical and sometimes magnetic fluctuations 
might have a large impact on the in-situ formation of new gas and dust species. The formation of one clump will have an effect on its close environment, more specifically on the nucleation and growth efficiency of new dust clumps. Dilution and multi-scattering effect will reduce the original contrast brightness of the clumps, but the continuous formation of new clumps combined with the effects of non-equilibrium drift-dependent dust formation will result in shell-like structures as observed on larger scales.  However, even within these shells, homogeneity will not prevail but relics from the original dust and gas clumps will remain. 
This complex envelope structure and local density enhancements should be incorporated into  chemical models that try to explain the formation of complex molecules in the envelope around IRC+10216 \citep[e.g.][]{Cordiner2009ApJ...697...68C,Agundez2010ApJ...724L.133A,Cherchneff2011A&A...526L..11C}.

\section{Conclusions} \label{Conclusions}

We have presented Herschel PACS observations at 70, 100, and 160\,$\mu$m that detect for the first time dust shells (or arcs) out to a radial distance of 320\arcsec\ in the envelope around the well-known carbon-rich AGB star IRC+10216. The shells are non-concentric and azimuthally incomplete. The shell/intershell density contrast is typically a factor of about four. Within an arc, local density variations are clearly visible. The arcs represent true density variations in the circumstellar envelope. 
From the PACS image, we have been able to infer the mass-loss history during the past 16\,000\,yr. It is argued that the nested shell structure should at least be visible out to the termination shock, at $\sim$390\arcsec. Instabilities in the turbulent astropause would wipe out any density contrast.

\begin{acknowledgements}
We are grateful to I.\ C.\ Le\~ao for providing the V-band image of IRC+10216. This research has made use of NASA's Astrophysics Data System Bibliographic Service and the SIMBAD database, operated at CDS, Strasbourg, France.
PACS has been developed by a consortium of institutes led by MPE (Germany) and including UVIE (Austria); KUL, CSL, IMEC (Belgium); CEA, OAMP (France); MPIA (Germany); IFSI, OAP/AOT, OAA/CAISMI, LENS, SISSA (Italy); IAC (Spain). This development has been supported by the funding agencies BMVIT (Austria), ESA-PRODEX (Belgium), CEA/CNES (France), DLR (Germany), ASI (Italy), and CICT/MCT (Spain). LD  acknowledges financial support from the Fund for Scientific Research - Flanders (FWO).  PR, BV, JB, and MG acknowledge support from the Belgian Federal Science Policy Office via the PRODEX Programme of ESA. FK acknowledges funding by the Austrian Science Fund FWF under project
number P23586-N16, RO under project number I163-N16.

\end{acknowledgements}
\vspace*{-1ex}
\vspace*{-.5cm}
\bibliographystyle{aa}
\bibliography{17360}

\begin{appendix}

\section{SPIRE 250\,$\mu$m image of IRC\,+10216}

The SPIRE 250\,$\mu$m was reduced using the same reduction technique as described in Sect.~\ref{Observations}. Only two faint arcs, at $\sim$1\arcmin\ and $\sim$2\arcmin\ are seen.

\begin{figure}[htp]
 \includegraphics[width=0.48\textwidth]{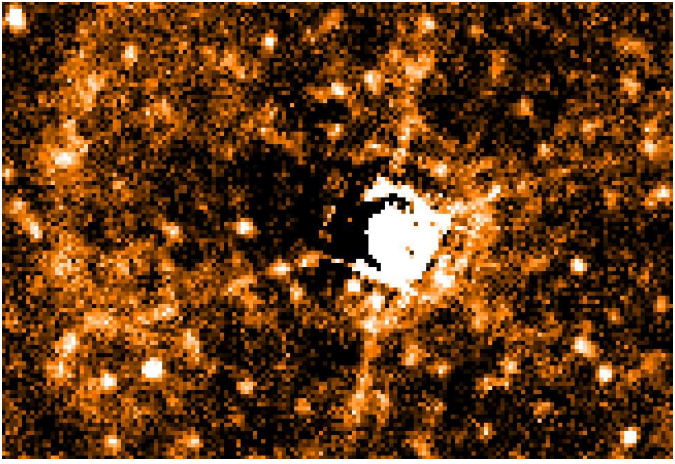}
\caption{SPIRE 250\,$\mu$m image of IRC\,+10216 after subtraction of the smooth halo of the circumstellar envelope. The field-of-view is 975\arcsec x 660\arcsec. North is at the top, east is to the left.}
\label{FIG_SPIRE250}
\end{figure}

\section{The turbulent interaction region with the ISM}\label{sec_bowshock}

\begin{figure}[htp]
 \includegraphics[width=0.45\textwidth]{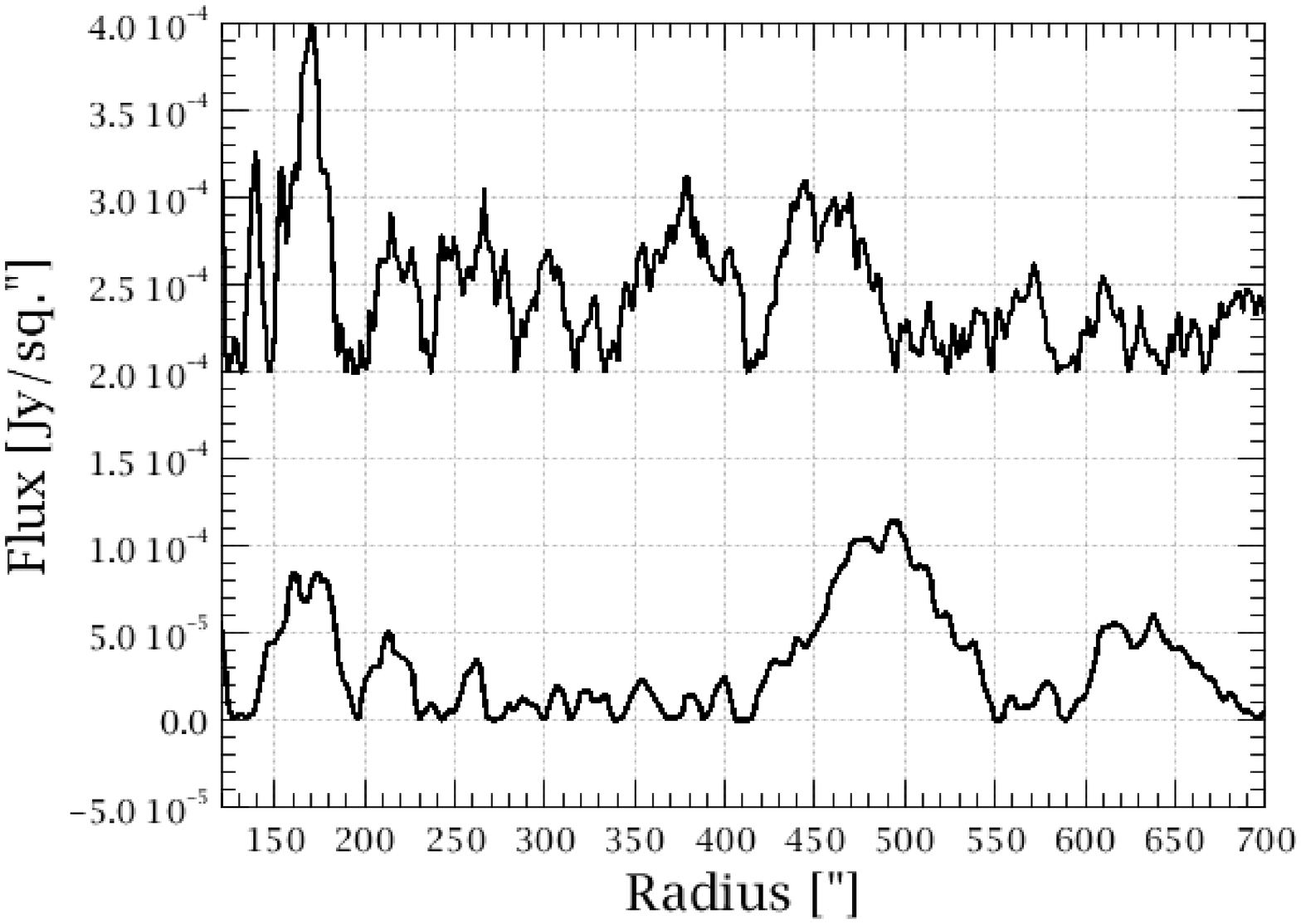}
\caption{Mean intensity enhancements in the PACS 100 (\textit{upper curve}) and 160\,$\mu$m (\textit{lower curve}) images in sector 5 (see Fig.~\ref{FIG_PACS100}), zooming into the bowshock region discussed by \citet{Ladjal2010A&A...518L.141L}. When measured from the central target, the bowshock is clearly detected at 480\arcsec\ in the 160\,$\mu$m image, and the corresponding feature is seen at 450\arcsec\ in the 100\,$\mu$m image. The feature around 370\arcsec\ in the 100\,$\mu$m image is not  an `arc', but more of a blob-like enhancement as can be seen in Fig.~\ref{FIG_PACS100}. }
\label{FIG_BOWSHOCK}
\end{figure}

The detection of the bowshock/astropause (see Sect.~\ref{Mdothistory}) around IRC+10216 using the Herschel PACS 160\,$\mu$m and SPIRE 250\,$\mu$m, 350\,$\mu$m, and 500\,$\mu$m images was published by \citet{Ladjal2010A&A...518L.141L}. To this detection, we now add the PACS 100\,$\mu$m image, while the bowshock is not visible in the 70\,$\mu$m image (see Fig.~\ref{FIG_PACS}). The bowshock is also visible in Fig.~\ref{FIG_PACS}  in the 100\,$\mu$m image situated closer to the central target than in the 160\,$\mu$m image. This is visualised in Fig.~\ref{FIG_BOWSHOCK}. At 100\,$\mu$m, the bowshock structure is at $\sim$450\arcsec\ from the central target, but it is at $\sim$480\arcsec\ in the 160\,$\mu$m image. From Fig.~2 in \citet{Ladjal2010A&A...518L.141L}, it seems that this trend is also seen for the longer wavelengths covered by SPIRE. This might indicate that smaller (and warmer) dust grains are responsible for the 100\,$\mu$m emission than at the longer wavelengths, in agreement with the theoretical predictions of \citet{vanMarle2011}.

\end{appendix}

\end{document}